\documentclass[a4paper]{jpconf}
\usepackage{color}
\usepackage{graphicx}

\def    \bta          {\mbox{\boldmath$\tau$}}

\begin{document}
\title{Induced magnetic moment in graphene with a nonmagnetic impurity}

\author{M Inglot$^{1}$ and V K Dugaev$^{1,2}$}

\address{$^1$Department of Physics, Rzesz\'ow University of Technology,
Al.~Powsta\'nc\'ow Warszawy 6, 35-959 Rzesz\'ow, Poland}
\address{$^2$Department of Physics and CFIF, Instituto Superior T\'ecnico,
Technical University of Lisbon, Av.~Rovisco Pais, 1049-001 Lisbon, Portugal}

\ead{ming@prz.edu.pl}

\begin{abstract}
We consider a two-dimensional crystalline monolayer of carbon
atoms with a single non-magnetic impurity. Using the Weyl
Hamiltonian to describe electronic energy spectrum near the Dirac
points, we calculate the wave function and energy of impurity
states, as well as the induced magnetic moment associated with
polarization of the electron system. We present a phase diagram of
the localized magnetic state as a function of the chemical
potential $\mu $ and  coupling constant $g_c$.
\end{abstract}

\section{Introduction}

Graphene has been attracting a lot of attention recently due to
its unusual electronic structure and, correspondigly, due to its
unique transport properties, which differ substantially from those
of traditional metals and semiconductors
\cite{novoselov05,geim07,castro09}. The key characteristic
features of graphene are the exact two-dimensionality of its
crystal structure and massless relativistic energy spectrum near
the Fermi surface. The Fermi surface transforms into isolated
points in the clean limit of undoped graphene, and the electronic
spectrum in the vicinity of these points can be described by the
relativistic Dirac model. It is now commonly believed that these
unusual properties of graphene make it an excellent candidate for
various device applications in nanoelectronics and/or spintronics.
\cite{avouris07}.

Although there is great interest in magnetic properties of
graphene, these are still not well understood. The main problem is
that the usual doping with magnetic impurities like Mn or Fe can
be ineffective because the magnetic atoms change dramatically the
structure of electronic states due to bonding with carbon atoms.
However, the possibility of doping with magnetic impurities, which
can lead to ferromagnetism in graphene, has been discussed in
several recent publications \cite{vozmediano05,peres05,dugaev06}.
The main interest now is related to the possibility of magnetism
induced by the presence of non-magnetic defects
\cite{yazyev07,uchoa08,yazyev08}.

In this work we consider formation of a local magnetic moment in
graphene, induced by a non-magnetic impurity. Assuming exchange
coupling between the localized spin and electronic states we
calculate magnetic polarization of the system and analyze
conditions for the appearance of  a local magnetic moment.

\section{Model of graphene with a single impurity}

The energy spectrum of graphene is described by the following
tight-binding Hamiltonian \cite{castro09},
\begin{eqnarray}
\mathcal{H}_{tb}=
\sum_i\varepsilon _Ac_{Ai}^{\dag}c_{Ai}
+\sum_n\varepsilon _B c_{Bn}^{\dag}c_{Bn}
+t\sum_{<i,n>}(c_{Ai}^{\dag}c_{Bn}+c_{Bn}^{\dag}c_{Ai})
\label{eq:1}
\end{eqnarray}
where the creation and annihilation operators correspond to
electrons localized in atomic sites of two nonequivalent
sublattices $A$ and $B$, with the hopping between the nearest
neighbors of different sublattices described by the parameter $t$.
The indices $i$ and $n$ refer to sites in the sublattices $A$ and
$B$, respectively. The onsite energy in  graphene is the same for
both sublattices, $\varepsilon _A=\varepsilon _B$.

By transforming Hamiltonian (1)  (with $\varepsilon _A=\varepsilon
_B$) to the momentum representation, one reproduces the energy
spectrum of graphene, which has two nonequivalent Dirac points in
the $k$ space, with a linear energy spectrum near these points.
The Dirac points are known to be the most interesting features of
the electronic structure, because the Fermi level of graphene is
usually located at these points or in their close vicinity.

In our calculations we restrict ourselves to the energy spectrum
near the Dirac points. The corresponding Hamiltonian is known in
quantum field theory  as the 2D Weyl Hamiltonian of massless
relativistic particles,
\begin{eqnarray}
\mathcal{H}_0=\left(
\begin{array}{cc}
0 & v\hbar \, (k_x-ik_y) \\
v\hbar \, (k_x+ik_y) & 0
\end{array}
\right) =v\hbar \bta \cdot {\bf k},
\label{eq:2}
\end{eqnarray}
where $v$ is the velocity of electrons, which in graphene is
$v\simeq 10^6\, m/s$, $\bta $ are the Pauli matrices acting in the
sublattice space (matrices of pseudospin), and we put the energy
onset in the Dirac points, $\varepsilon _A=\varepsilon _B=0$.

\begin{figure}[h]
    \centering
        \includegraphics[width=0.25\textwidth]{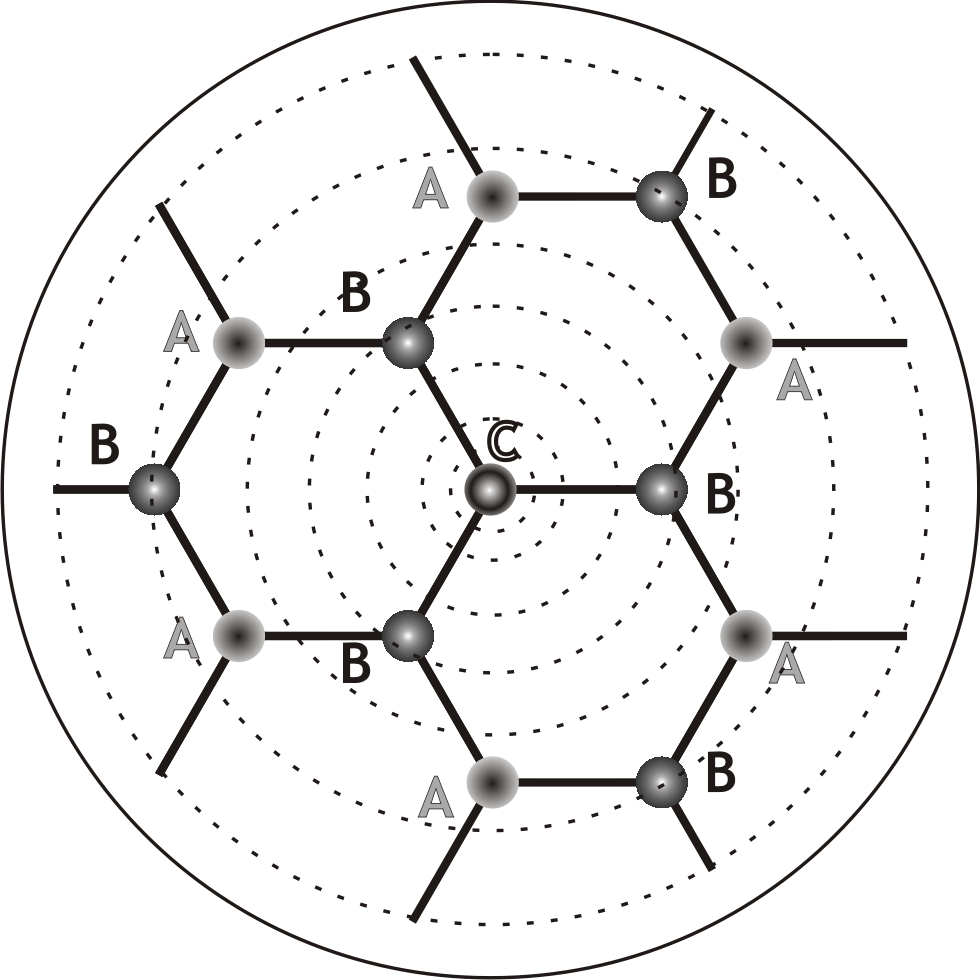}
    \caption{Schematic view of an impurity $C$, which substitutes a carbon atom in the sublattice $A$ of graphene.}
    \label{fig:zab2}
\end{figure}

In the following we assume that one carbon atom in the sublattice
$A$ is replaced by an impurity atom $C$ (located at
$\mathbf{r}=\mathbf{0}$), as shown schematically in Fig.~1. This
replacement can be formally included in the Hamiltonian (1) as a
term with onsite energy $\varepsilon _C$ at the site $i=0$, so
that the perturbation is $V_0=(\varepsilon _C-\varepsilon
_A)c_{A0}^{\dag}c_{A0}$. After reducing to the Weyl Hamiltonian,
the corresponding perturbation term in the polar coordinates
($\rho, \phi$)
\begin{equation}
    \mathcal{H}_{int}=\left( \begin{array}{cc}
    V_0\, \delta (\rho ) & 0 \\
    0 & 0
    \end{array} \right) .
    \label{eq:3}
\end{equation}

\section{Localized electronic states}

Our task now is to find impurity states described by the
Hamiltonian $\mathcal{H}=\mathcal{H}_0+\mathcal{H}_{int}$, with
$\mathcal{H}_0$ and $\mathcal{H}_{int}$ defined by Eqs.~(2) and
(3), respectively. The corresponding Schr\"odinger equation,
$(\mathcal{H}-\varepsilon )\Psi (\rho,\varphi )=0$, gives us the
energy and eigenfunction  of the localized electronic states.

Solving the Schr\"odinger equation for $\rho \ne 0$ we find a
general solution in the form of a superposition of the Bessel
functions $J_m(\xi )$ and $Y_m(\xi )$, where $\xi =\rho
\varepsilon /\hbar v)$ and $m$ is the magnetic quantum number. It
turns out that only the following pseudospinor form of the
solution gives us the localized states near the Dirac point:
\begin{eqnarray}\label{Ffal2}
    \Psi _m(\varrho ,\varphi )
    =N \left( \begin{array}{l}
    e^{im\varphi}\,Y_m(\xi ) \\
    \displaystyle{i\, e^{i(m+1)\varphi}}\, Y_{m+1}(\xi )
    \end{array} \right) ,
    \label{eq:4}
\end{eqnarray}
where $N$ is the normalization factor.

Energy of the localized states can be found by integrating the
Schr\"odinger equation with the wavefunction (4) over a small area
near the point $\rho =0$. Within the lattice model, this
corresponds to integration over $\rho <a_0$, where $a_0$ is the
lattice parameter. This way we account for the boundary condition
at $\rho =0$ with the $\delta $-like perturbation localized at
this point. The ground state corresponds to the choice $m=0$ or
$m=-1$. Both states give us the same energy of the localized
state,
\begin{equation}
\varepsilon _0
=\frac{2\pi\,\hbar ^2v^2}
{V_0\ln \left( a_0 |\varepsilon _0|/\hbar v\right) }
\simeq
\frac{2\pi\,\hbar ^2v^2}
{V_0\ln \left( 2\pi a_0\hbar v/|V_0|\right) }.
\label{eq:5}
\end{equation}
As we see from the solution (5), the energy of the localized state
$\varepsilon _0\to 0$ as the perturbation $|V_0|$ increases. For
the parameters typical of graphene, the sign of \ logarithm in (5)
is negative, and therefore the sign of the energy state
$\varepsilon _0$ is opposite to that of perturbation potential. In
other words, the attractive (negative) potential, $V_0<0$, gives
us a state with the energy $\varepsilon _0>0$ and {\it vice
versa}. This is just the opposite behavior as compared to the
impurity state in traditional semiconductors -- if we assume that
graphene is similar to the gapless semiconductors. An important
point is the above mentioned degeneracy of localized states, which
is related to the pseudospinor wave function in graphene.

We have also checked the result (5) for the energy of impurity
state by calculating the exact $T$-matrix of scattering from the
impurity described by the matrix perturbation (3). Both
calculation methods give exactly the same results.

\section{Induced magnetic moment}

The impurity state in graphene is occupied by an electron when the
Fermi level is above the impurity level (5). Hence, in our
consideration  we assume that the location of the Fermi level is
an independent parameter which is not related with the single
impurity state under consideration. In reality, this assumption is
well justified  because position of the Fermi level in graphene is
usually related with various defects or simply can be controlled
by a gate voltage.

The single electron localized at the impurity has the magnetic
moment $\mu _B$. In the following consideration we assume that the
Coulomb interaction does not allow for two or more electrons to
occupy the same impurity state, which means that the Hubbard
energy is sufficiently large.

Magnetic moment of the localized electron polarizes the electron
system, inducing magnetic moment whose magnitude depends on the
magnetic polarizability of the electron gas in graphene. Assuming
exchange coupling of electrons in graphene with the localized
electrons we calculate the induced magnetic moment. The
Hamiltonian of such an exchange interaction is
\begin{eqnarray}
H^m_{int}=\frac12\, g_c\mu_z(\mathbf{r})\sigma_z,
\label{eq:6}
\end{eqnarray}
where $g_c$ is the coupling constant, $\mu_z({\bf r})$ is the
spatial distribution of magnetization associated with the
wavefunction profile of the localized electron in Eq.~(4),
\begin{eqnarray}\label{MomentMag}
\mu_z({\bf r})=g\mu_B|\Psi({\bf r})|^2,
\label{eq:7}
\end{eqnarray}
and $\sigma_z$ is the Pauli matrix for electron spin. We take the
quantization axis along the spin orientation of the localized
electron.

The induced magnetic moment ${\bf M}({\bf r})$ of electrons in
graphene can be calculated using the quantum field theoretical
method, used for calculation of magnetic polarizability
\cite{agd}. Accordingly, one can present it as a loop Feynman
diagram, which gives
\begin{eqnarray}
M_z({\bf r})=ig_cg\mu_B\,{\rm Tr}\int
d^2\mathbf{r'}\,
\int \frac{d\varepsilon }{2\pi}\,\sigma_z\,G_0
\left(\varepsilon ,\mathbf{r}-\mathbf{r'}\right)
\sigma_z G_0\left(\varepsilon ,{\bf r'}-{\bf r}\right)\, \mu_z({\bf r'}),
\label{eq:8}
\end{eqnarray}
where $G_0(\varepsilon, {\bf r})$ is the Green function of
electrons in graphene in the energy-coordinate representation. The
calculation of the Green function using the Hamiltonian (2) gives
\begin{eqnarray}
G_0(\varepsilon ,\pm \mathbf{r})
=-\frac{i\varepsilon }{4\hbar ^2v^2}
H_0^{(1)}\left(\frac{r\varepsilon }{\hbar v}\right)
\pm \bta \cdot {\bf r}\frac{\varepsilon }{4r\hbar ^2v^2}
H_1^{(1)}\left(\frac{r\varepsilon }{\hbar v}\right) ,
\label{eq:9}
\end{eqnarray}
where $H^{(1)}_{0,1}$ are the Hankel functions \cite{abramowitz}.
Using Eqs.~(7)-(9) we find
\begin{eqnarray}
M_z(\mathbf{r})
=-\frac{g_cg\mu_B\zeta }{4\pi \hbar v}\int d^2\mathbf{r'}\;
\frac{\mu_z(\mathbf{r'})}{|{\bf r}-{\bf r}'|^3} ,
\label{10}
\end{eqnarray}
where
\begin{eqnarray}
\zeta =\int\limits^{\infty}_{0}d\xi \,
\xi^2\left[ K^{2}_0(\xi)+K^{2}_1(\xi)\right] ,
\label{eq:11}
\end{eqnarray}
and
\begin{eqnarray}
K_m(\xi)=\frac{i^{m+1}\pi}{2}H_m^{(1)}(i\xi)
\label{eq:12}
\end{eqnarray}
is the the modified Bessel function \cite{abramowitz}.

\begin{figure}[t]
    \centering
        \includegraphics[width=0.5\textwidth]{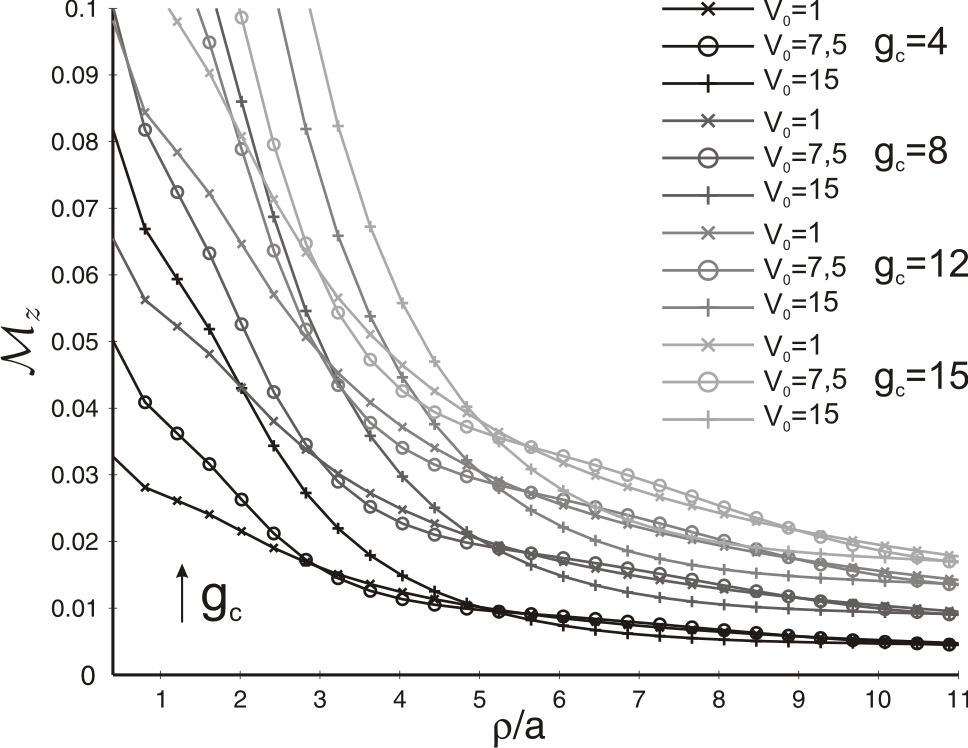}
    \caption{The density of induced magnetic moment as a function of the distance $\rho $ for different
    coupling parameters.}
    \label{fig:Mag1a}
\end{figure}

Using Eqs.~(10)-(12) one can write  the induced magnetic moment in
the following form:
\begin{eqnarray}
M_z(\rho)
=\frac{g_cg^2\mu_B^2\pi}{8\hbar v}
\int _0^\infty
\frac{\rho' |\psi(\rho ')|^2 \mathcal{E}
\left( \varsigma\right) }
{(\rho +\rho ')(\rho -\rho ')^2}\; d\rho '
\label{eq:13}
\end{eqnarray}
where $\mathcal{E}\left( \varsigma\right)$ is the elliptic integral of the second kind and $\varsigma=2\sqrt{\rho \rho'}/ {(\rho+\rho')}$. This expression was used for the numerical calculations, and the results are presented in Fig.~2.

We have also calculated the total induced magnetic moment
\begin{eqnarray}
M_0(V_0,g_c)=2\pi \int\limits_0^\infty M_z(\rho )\, \rho d\rho
\label{eq:14}
\end{eqnarray}
The dependence of $M_0$ on the strength of impurity potential
$|V_0|$ is found to be rather weak, whereas the main factor which
may enhance $M_0$ is the magnitude of the coupling parameter
$g_c$.

\section{Magnetic coupling of the localized spin with induced magnetic moment}

The magnetic interaction of the localized spin with induced
magnetic density $M_z(\rho )$ leads to the renormalization of the
impurity energy level \cite{abrikosov73}. This interaction can be
written  as
\begin{eqnarray}
E_{int}=-2\pi g_c\int\limits_0^{\infty}M_z(\mathbf{\varrho})
|\Psi(\varrho)|^2\, \rho \, d\rho .
\label{eq:15}
\end{eqnarray}
We have calculated numerically the interaction (15) and found the
corresponding shift of the energy level. Our results are presented
in Fig.~3.

\begin{figure}[h]
    \centering
        \includegraphics[width=0.6\textwidth]{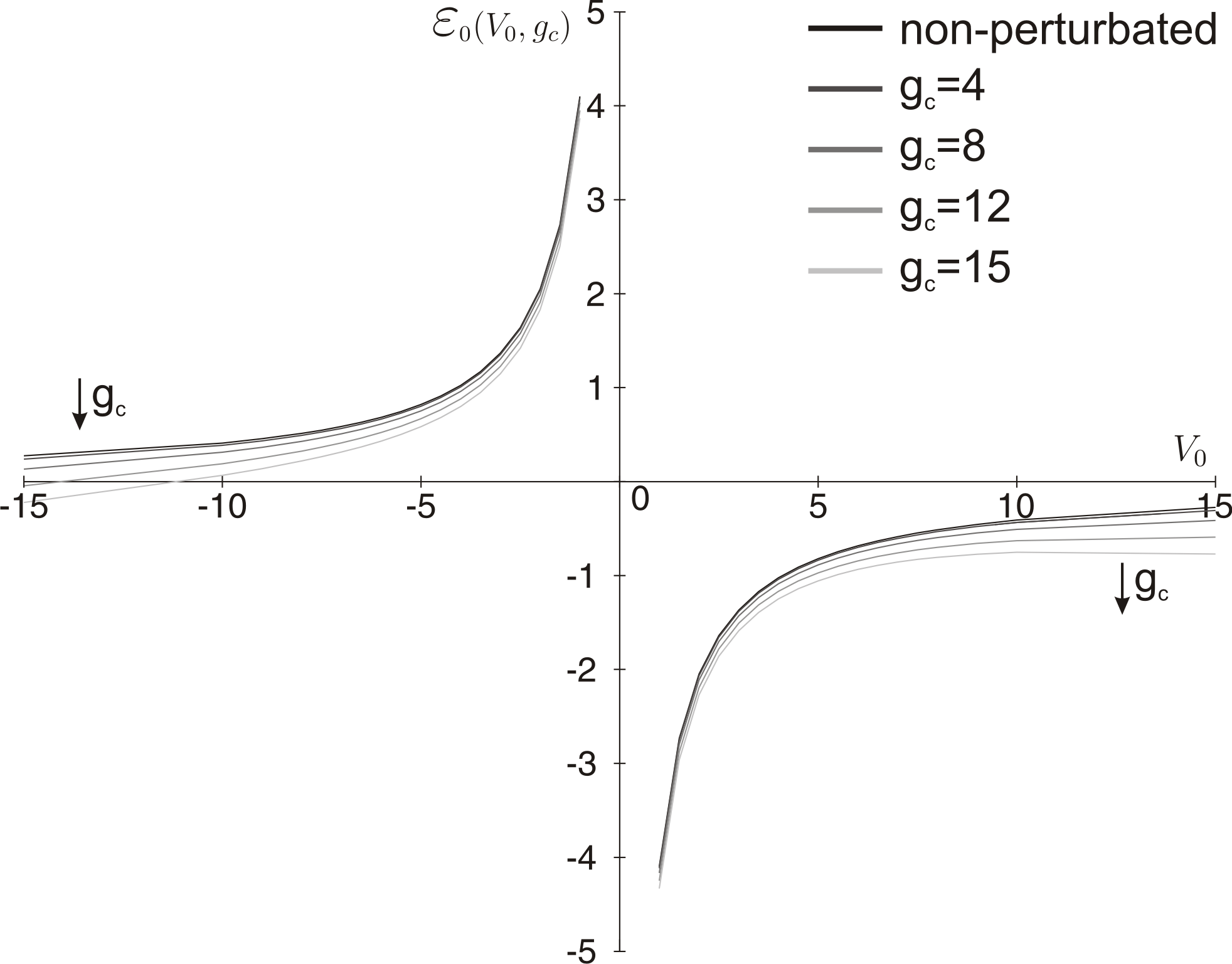}
    \caption{Energy level of the localized electron interacting with the induced magnetic moment for different values of the coupling constant $g_c$. The energy in the units of hopping t.}
    \label{fig:Eint1}
\end{figure}

\section{Phase diagram}

The obtained results can be presented in the form of a phase
diagram, Fig.~4, which shows the regions of the parameters $V_0$
and $\mu $, for which the impurity level is occupied with electron
inducing the local magnetization around impurity. It should be
pointed out that the level can be occupied even if the
nonrenormalized position of the level is above the Fermi energy
\cite{abrikosov73}. This result is related to the effect of onsite
magnetic correlations.

\begin{figure}[t]
    \centering
        \includegraphics[width=0.45\textwidth]{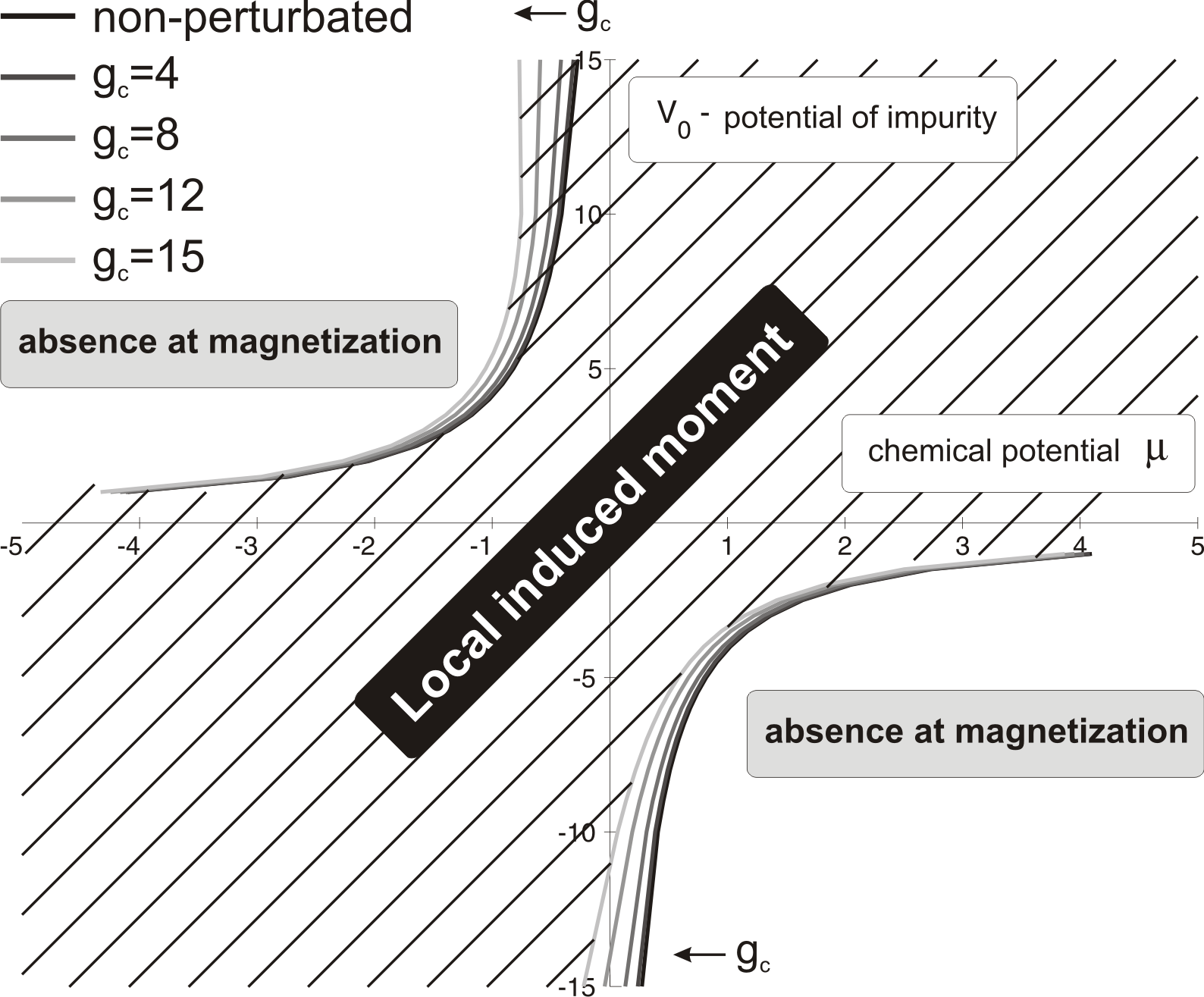}
    \caption{The phase diagram with the dashed area corresponding to the local magnetic moment induced by a
     nonmagnetic impurity.}
    \label{fig:diagfazowy}
\end{figure}

\section{Conclusion}

The results of our calculations show that the nonmagnetic impurity
can create an induced localized magnetic state due to the local electronic correlations.
It should be emphasized that this effect is not related to free electrons because
the density of free electrons in graphene can be vanishingly small.
However, the magnetic polarizability
of the electron system is large and long-range even without free electrons, which is
the main reason for the local ferromagnetism.

\ack
This work is supported by the Polish Ministry of
Science and Higher Education as a research project in years 2007 -- 2010
and by FCT Grant PTDC/FIS/70843/2006 in Portugal.

\end{document}